\documentclass[prd,twocolumn,superscriptaddress,showpacs,nofootinbib,preprintnumbers]{revtex4}

\usepackage{amsmath}
\usepackage{amsfonts}
\usepackage{graphicx}
\usepackage{dcolumn}
\usepackage{hyperref}

\def\be{\begin{equation}}
\def\ee{\end{equation}}
\def\bea{\begin{eqnarray}}
\def\eea{\end{eqnarray}}

\def\nn{\nonumber \\}
\def\e{{\rm e}}
\def\d{{\rm d}}

\def\c{\hspace{-5pt}}

\def\5{\overline 5}

\begin{document}

\title{Properties of singularities in (phantom) dark energy universe}

\author{Shin'ichi Nojiri}
\email{snojiri@yukawa.kyoto-u.ac.jp}
\affiliation{Department of Applied Physics,
National Defence Academy,
Hashirimizu Yokosuka 239-8686, Japan}
\author{Sergei D.~Odintsov}
\email{odintsov@aliga.ieec.uab.es}
\affiliation{Instituci\`o Catalana de Recerca i Estudis
Avan\c{c}ats (ICREA)  and Institut d'Estudis Espacials de Catalunya (IEEC),
Edifici Nexus, Gran Capit\`a 2-4, 08034 Barcelona, Spain}
\author{Shinji Tsujikawa}
\email{shinji@nat.gunma-ct.ac.jp}
\affiliation{Department of Physics, Gunma National College of
Technology, Gunma 371-8530, Japan}

\date{\today}

\vskip 1pc
\begin{abstract}
The properties of future singularities are investigated in the universe 
dominated
by dark energy including the phantom-type fluid.
We classify the finite-time singularities into four classes and explicitly 
present
the models which give rise to these singularities by assuming the form of
the equation of state of dark energy.
We show the existence of a stable fixed point with an equation of state
$w<-1$ and numerically confirm that this is actually a late-time attractor
in the phantom-dominated universe.
We also construct a phantom dark energy scenario coupled to dark matter
that reproduces singular behaviors of the Big Rip type
for the energy density and the curvature of the universe.
The effect of quantum corrections coming from
conformal anomaly can be important when the curvature grows large,
which typically moderates the finite-time singularities.
\end{abstract}


\maketitle
\vskip 1pc

\section{Introduction}

The increasing evidence from the observational data indicates
that  (linear) equation of state (EOS) parameter $w$ lies
in a narrow strip around $w=-1$ quite likely being below of
this value \cite{obser}.
The region where the EOS parameter $w$
is less than $-1$ is typically referred as a phantom dark energy universe.
This is caused by the fact that when the phantom EOS is constructed in 
terms
of a scalar field the corresponding kinetic term is chosen to have a wrong
sign (negative kinetic energy). Of course, this is not the only
possibility: the phantom-like value for $w$ may appear from
Brans-Dicke (BD) scalar-tensor gravity, from non-standard (negative) 
potentials,
from the non-minimal coupling of scalar Lagrangian with gravity
or even usual matter may appear in phantom-like form.
Recent works \cite{phantom,phantom1,ENO} have been
devoted to the study of phantom cosmologies produced by different models.

The existence of the region with $w<-1$
(if such a phase in the universe evolution indeed occurs)
opens up a number of fundamental questions.
For instance, the entropy of such universe is negative
(or the characteristic temperatures should be
negative). The dominant energy condition (DEC) for phantom matter is
violated, as a rule. The phantom dominated universe ends up with a
finite-time future singularity called Big Rip or Cosmic Doomsday
(see Refs.~\cite{CKW,final}).
This last property attracted much attention
and brought the number of speculations up to the explicit calculation
of the rest life-time of our universe!
In its turn, such a study
motivated the mathematical investigation of singularities when the DEC
or the strong energy condition (SEC)
is violated \cite{Barrow} (see also Ref.~\cite{Barrow2}).
However, the Big Rip singularity is characterized by the growth of the 
energy
and curvature invariants with a divergent scale factor
at Big Rip time. The energy scale may grow up to
Planck one, giving rise to the second Quantum Gravity era.
Eventually, quantum effects become important near the singularity
where they may moderate or even prevent the singularity
\cite{quantum,final}.

In such circumstances, it is fundamentally important to understand
the properties of singularities in phantom dark energy universe,
to classify them and to search for realistic ways to avoid the
singularities. The present article is devoted to the investigation of
the above circle of problems. We study phenomenological
models in which the pressure density $p$ of dark energy is
given in terms of the function of the energy density
$\rho$, i.e., $p=-\rho-f(\rho)$.
When the function $f(\rho)$ is zero, the EOS parameter
$w$ is equal to $-1$ (cosmological constant EOS).
We classify the types of future singularities
and show explicit examples of phantom dark energy which realize
all mentioned types of singularities. The evolution
of such dark energy universe at late times (near to singularity) is
investigated both analytically and numerically.

The paper is organized as follows.
The general system of dark energy coupled with dark matter
is considered in Sec.~II.
The corresponding background equations are written in autonomous
form which permits to analyze the fixed points and
attractor solutions. We also study the stability of fixed points against
perturbations and show the existence of a stable critical point with
$w<-1$.  Sec.~III is devoted to the classification
of finite-time future singularities (four different types) and their
explicit realization in terms of the universe with the
equation of state: $p=-\rho-f(\rho)$.
We present the existence of singularities of much less explicit type than 
the
standard Big Rip where the scale factor is finite at
Rip time (including the type given in Ref.~\cite{Barrow}).
We construct examples of phantom dark energy universes in which all
four types of singularities appear.
The violation of strong and dominant energy conditions is investigated as
well. We also present one example of the model which admits the
transition from $w<-1$ to $w>-1$.
In Sec.~IV the relation between the EOS function $f(\rho)$ and the
appearance of singularities is studied.
Sec.~V is devoted to the numerical study
of attractor solutions in dark energy models with non-relativistic dark 
matter.
The phase plane analysis confirms the structure of singularities and
the stability around the fixed points.

In Sec.~VI we present coupled phantom/fluid dark matter models
for several specific coupling functions. These models generalize some of 
the
examples of the previous sections. We explicitly show multiple scalar field
models that exhibit the similar singular structure as in the coupled 
phantom
scenario. In Sec.~VII the role of quantum effects
is investigated for the types of singularities classified in Sec.~III.
In accordance with previous attempts in this direction
\cite{quantum,final,sri} it is demonstrated that quantum effects can
moderate the finite-time singularities.
Summary and some outlook are given in the final section.

\section{Autonomous system \label{SII}}

Let us consider a system with two fluids: (i) dark energy with
an equation of state: $p=p(\rho)$, and (ii) a barotropic perfect fluid
with an equation of state: $p=w_m \rho$.
We wish to study the case of a nonrelativistic
dark matter ($w_m=0$), but we keep our discussion in
general for the moment.
In a spatially flat Friedmann-Lemaitre-Robertson-Walker (FLRW)
metric with a scale factor $a$, the background equations
are given by
\bea
\label{rhoeq}
& & \dot{\rho}+3H(\rho+p)= -Q,  \\
\label{rhoeq2}
& & \dot{\rho}_m+3H(\rho_m+p_m)= Q, \\
\label{Heq}
& & \dot{H}=-\frac{\kappa^2}{2}
(\rho+p+\rho_m+p_m)\,,
\eea
where $H \equiv \dot{a}/a$ is the Hubble rate, and
$\kappa^2 \equiv 8\pi G$ with Newton's gravitational constant $G$.
Here a dot denotes the derivative with respect to cosmic time $t$.
We accounted for the interaction term $Q$ between dark energy and
the barotropic fluid.
We also have the constraint equation for the Hubble rate:
\bea
\label{FRWH}
H^2=\frac{\kappa^2}{3}(\rho+\rho_m)\,.
\eea

In analogy with scalar field dark energy models, we introduce
the following ``kinematic'' and ``potential'' terms:
\bea
\rho_K \equiv (\rho+p)/2\,,~~~
\rho_V \equiv (\rho-p)/2\,.
\eea
For scalar fields one has $\rho=\frac12 \epsilon \phi^2+V(\phi)$
and $p=\frac12 \epsilon \phi^2-V(\phi)$
(a normal field corresponds to $\epsilon=1$ and
a phantom to $\epsilon=-1$).
We shall introduce the following dimensionless quantities:
\bea
x \equiv \frac{\kappa^2 \rho_K}{3H^2}\,,~~~~
y \equiv \frac{\kappa^2 \rho_V}{3H^2}\,.
\eea
Then the above background equations can be written
in an autonomous form as an extension of Ref.~\cite{CLW}:
\bea
\frac{\d x}{\d N}&=&-\left[1+p'(\rho) \right]
\left[3+\frac{Q}{2H\rho_K}\right]x \nn
&& +3x \left[2x+(1+w_m)(1-x-y)\right]\,, \\
\frac{\d y}{\d N}&=&-\left[1-p'(\rho) \right]
\left[3x+\frac{Q}{2H\rho_V}y \right] \nn
&& +3y \left[2x+(1+w_m)(1-x-y)\right]\,, \\
\frac{1}{H} \frac{\d H}{\d N}&=&
-\left[3x+\frac32 (1+w_m)(1-x-y) \right]\,,
\eea
together with the constraint
\bea
\Omega_m \equiv \frac{\kappa^2 \rho_m}{3H^2}
=1-x-y \,,
\eea
where $N \equiv {\rm ln}\, a$ and
$p'(\rho)$ is a function of $\rho$
defined as $p'(\rho) \equiv \d p/\d \rho$.
Since $\kappa^2 \rho_m/(3H^2) \ge 0$, the parameter
range of $x$ and $y$ is restricted to be
\bea
\label{xyres}
x+y \le 1\,.
\eea

When $w_m=0$ and $Q=0$, these are simplified as
\bea
\label{dx}
& & \frac{\d x}{\d N} = 3x \left[x-y-p'(\rho) \right]\,, \\
\label{dy}
& & \frac{\d y}{\d N} = -3\left[1-p'(\rho) \right]x
+3y \left[1+x-y\right]\,, \\
& &\frac{1}{H} \frac{\d H}{\d N}=
-\frac32 (1+x-y)\,.
\eea
Let us consider a situation in which $p'(\rho)$
asymptotically approaches a constant, i.e., $p'(\rho) \to w$.
In this case the equation of state for dark energy is
\bea
\label{weq}
w \equiv \frac{p}{\rho}=\frac{x-y}{x+y}\,.
\eea
In the phase plane in terms of $x$ and $y$, the trajectory
corresponding to the constant $w$ is a straight line, i.e.,
\bea
\label{yandx}
y=\frac{1-w}{1+w}x\,.
\eea

Setting $\d x/\d N=0$ and $\d y/\d N=0$ in
Eqs.~(\ref{dx}) and (\ref{dy}),
one obtains the following fixed points:
(i) $(x, y)=(0, 0)$, (ii) $(x, y)=[(1+w)/2, (1-w)/2]$ 
and (iii) $(x, y)=(0, 1)$.
The fixed point (i) corresponds to the barotropic fluid
dominant solution ($\Omega_m \to 1$).
The point (ii) is the dark energy dominant
solution, that is, $\Omega_{\rm DE} \equiv x+y \to 1$.
In the case (ii) we have
\bea
\frac{1}{H} \frac{\d H}{\d N}=
-\frac32 (1+w)\,,
\eea
which means that the Hubble rate
increases for $w<-1$.
The class (iii) corresponds to the potential-dominant solution 
with $\Omega_{\rm DE} \equiv x+y \to 1$.

One can investigate the stability of the system
by considering small perturbations $\delta x$
and $\delta y$ around the fixed point $(x_0, y_0)$,
i.e., $x=x_0+\delta x$ and
$y=y_0+\delta y$. By Eqs.~(\ref{dx}) and
(\ref{dy}), we obtain the linearized equations
\bea
& & \hspace*{-1.5em} \frac{\d }{\d N} \delta x =
3(2x_0-y_0-w)\delta x-3x_0 \delta y\,, \\
& & \hspace*{-1.5em} \frac{\d }{\d N} \delta y =
3(-1+y_0+w)\delta x+3(1+x_0-2y_0)\delta y\,,
\eea
This can be written by using a matrix ${\cal M}$:
\bea
& &
\frac{\d}{\d N}
\left(
\begin{array}{c}
\delta x \\
\delta y
\end{array}
\right) = {\cal M} \left(
\begin{array}{c}
\delta x \\
\delta y
\end{array}
\right) \,.
\label{uvdif}
\eea

One can study the stability of critical points
against perturbations by evaluating the eigenvalues of the
matrix ${\cal M}$.
For the class (i) one has $\lambda=-3w$ and $\lambda=3$,
which means that the solution is an unstable node for $w<0$
and a saddle point for $w>0$.
For the class (ii) we obtain $\lambda=3w$ and
$\lambda=3(w+1)$.
Therefore the solution is a stable node for $w<-1$,
a saddle point for $-1<w<0$, and an unstable node
for $w>0$.
Then only the stable fixed point is the class (ii) with
$w<-1$. 
The class (iii) corresponds to the eigenvalues:
$\lambda=-3(1+w)$ and $\lambda=-3$.
Since $w=-1$ in this case by Eq.~(\ref{weq}), one 
has $\lambda=0, -3$. Therefore the fixed point (iii)
is marginally stable.

The above argument corresponds to the one in which
$p'(\rho)$ asymptotically approaches a constant $w$.
We can expect that this argument may be applied to
the case with a dynamically changing $p'(\rho)$
by following ``instantaneous'' fixed points as
in Refs.~\cite{NNR,CGST}.
We will check this behavior numerically
in Sec.~V.

\section{Models of Future Singularities \label{SIII}}

In the present section we will consider the dark energy universe
models which contain finite-time, future singularities.
It is clear that depending on the content of the model
such singularities may behave in different ways.
That is why it is useful to classify the future singularities
in the following way:
\begin{itemize}
\item  Type I (``Big Rip'') : For $t \to t_s$, $a \to \infty$,
$\rho \to \infty$ and $|p| \to \infty$
\item  Type II (``sudden'') : For $t \to t_s$, $a \to a_s$,
$\rho \to \rho_s$ and $|p| \to \infty$
\item  Type III : For $t \to t_s$, $a \to a_s$,
$\rho \to \infty$ and $|p| \to \infty$
\item  Type IV : For $t \to t_s$, $a \to a_s$,
$\rho \to 0$, $|p| \to 0$ and higher derivatives of $H$ diverge.
\end{itemize}
Here $t_s$, $a_s$ and $\rho_s$ are constants with
 $a_s\neq 0$.
The type I is so-called the Big Rip singularity \cite{CKW}
which emerges for the phantom-like equation of state: $w<-1$.
The type II corresponds to the sudden future 
singularity in \cite{Barrow} at which 
$a$ and $\rho$ are finite but $p$ diverges.
The type III appears for the model with $p=-\rho-A \rho^{\alpha}$
\cite{Stefancic}, which is
different from the sudden future singularity 
in the sense that $\rho$ diverges.
This type of singularity has been discovered in the model
of Ref.~\cite{final} where the corresponding Lagrangian
model of a scalar field with potential has been constructed.
The type IV is a new type of singularity which appears in the model
described below.

In this section we consider the dark energy EOS
characterized by
\be
\label{EOS1}
p=-\rho - f(\rho)\,,
\ee
where $f(\rho)$ can be an arbitrary function in general.
Note that such EOS maybe equivalent to bulk viscosity, 
see Refs.~\cite{Barrrowvis}.
The function $f(\rho)\propto \rho^\alpha$ with a constant $\alpha$
was proposed in Ref.~\cite{final}
and was investigated in detail in Ref.~\cite{Stefancic}.
When $Q=0$ in Eq.~(\ref{rhoeq}),
the scale factor is given by
\be
\label{EOS4}
a=a_0\exp \left(\frac{1}{3}\int \frac{\d\rho}{f(\rho)}\right)\,,
\ee
where $a_0$ is a constant.
Here it is assumed $f(\rho)$ does not vanish
for all values of $\rho$.
When $f(\rho)=0$ everywhere, the standard $\Lambda$CDM cosmology is
recovered, so the above choice of EOS is a sound way to study the
deviations from such cosmology.

In this section we do not implement the contribution of a barotropic
fluid, i.e., $\rho_m=0$, $p_m=0$ and $Q=0$
in Eqs.~(\ref{rhoeq})-(\ref{FRWH}).
Combining Eqs.~(\ref{rhoeq}) and (\ref{FRWH}) with (\ref{EOS1}),
one finds
\be
\label{tint}
t=\int \frac{\d \rho}{\kappa \sqrt{3\rho} f(\rho)}\,,
\ee
which is used  later.
Eqs.~(\ref{Heq}) and (\ref{FRWH}) give
\be
\label{EOS5}
\frac{\ddot{a}}{a}=-\frac{\kappa^2}{6}\left(\rho + 3p\right)
=\frac{\kappa^2}{6}\left[2\rho + 3f(\rho)\right]\ .
\ee
For example, for the specific choice of EOS:
$f(\rho)=-2\rho/3 + \rho_0 \sin (\rho_1 /\rho)$,
with constants $\rho_0$ and $\rho_1$, the universe iterates the transition 
between the
acceleration ($\ddot{a}>0$) and deceleration ($\ddot{a}<0$).

\subsection{The model of transition from $w>-1$ to $w<-1$}

In this subsection we study the model in which
EOS changes from $w>-1$ to $w<-1$.

When $w$ is a constant, one has
$a\sim t^{{2}/{3(1+w)}}$ and $H\sim {2}/{3(1+w)}t$ for $w>-1$,
and $a\sim \left(t_s - t\right)^{{2}/{3(1+w)}}$ or
$H\sim -{2}/{3(1+w)}(t_s - t)$ for $w<-1$.
It is interesting to understand what kind of $f(\rho)$
admits the transition between the region with $w>-1$
and that with $w<-1$.
Let us consider the following simple model
\be
\label{EOS7}
a(t)=a_0 \left(\frac{t}{t_s - t}\right)^n \,.
\ee
Here $n$ is a positive constant and  $0<t<t_s$.
The scale factor diverges with a finite time ($t \to t_s$)
as in the Big Rip singularity.
Therefore $t_s$ corresponds to the life time of the universe.
When $t \ll t_s$, $a(t)$ evolves as $t^n$,
which means that the effective EOS
is given by $w=-1 + 2/(3n)>-1$.
On the other hand, when $t \sim t_s$, it appears $w=-1 - 2/(3n)<-1$.

{}From Eq.~(\ref{EOS7}) the Hubble rate is given by
\be
\label{EOS8}
H=n\left(\frac{1}{t} + \frac{1}{t_s - t}\right)\,.
\ee
Then using Eq.~(\ref{FRWH}), we find
\be
\label{EOS10}
\rho=\frac{3n^2}{\kappa^2}
\left(\frac{1}{t} + \frac{1}{t_s - t}\right)^2\,.
\ee
Hence, both $H$ and $\rho$ have minima at $t=t_s/2$
with the values
\be
\label{EOS11}
H_{\rm min}=\frac{4n}{t_s}\ ,\quad
\rho_{\rm min}=\frac{48n^2}{\kappa^2 t_s^2}\ .
\ee

By deleting $t$ in $\dot{\rho}$,
one obtains
\be
\label{EOS12}
\dot{\rho} =\pm 2\rho \left\{ \frac{\rho\kappa^2}{3n^2}
- \frac{4}{n t_s}\left(\frac{\kappa^2 \rho}{3}
\right)^{\frac{1}{2}} \right\}^{\frac{1}{2}}\ .
\ee
Here the plus sign in Eq.~(\ref{EOS12}) corresponds to 
the region $t>t_s/2$ and the minus one to
the region $t<t_s/2$.
Combining Eq.~(\ref{EOS12}) with Eq.~(\ref{rhoeq}),
it follows
\be
\label{EOS14}
f(\rho) = \pm \frac{2\rho}{3n}\left\{ 1 -
\frac{4n}{t_s}\left(\frac{3}{\kappa^2\rho}
\right)^{\frac{1}{2}}\right\}^{\frac{1}{2}}\,.
\ee
Therefore the EOS needs to be double-valued in order for
the transition to occur between the region $w<-1$ and the region $w>-1$.
Such a double-valued EOS
often appears when there is a first-order phase transition.
As we will see later in Sec.~VI, if there are two kinds of matter or 
energy,
a value of $\rho$ can correspond to several values of $p$.
We should also note $f(\rho_{\rm min})=0$, that is, the minima of $\rho$ 
and $H$
correspond to the transition point: $w=-1$.

The singularity at $t=t_s$ corresponds to the Big Rip type
characterized by $a \to \infty$, $\rho \to \infty$ and $|p| \to \infty$
for $t \to t_s$. In this region $f(\rho)$ behaves as $f(\rho) \sim
2\rho/3n$ due to Eq.~(\ref{EOS14}), which means that the pressure
$p$ is linear in $\rho$, i.e., $p=-\rho-2\rho/3n$.
Therefore this gives the constant EOS: $w=\rho/p=-1-2/3n$.
In another asymptotic limit $t \to 0$,
one also obtains the constant value of $w$, i.e., $w=-1+2/3n$.

For the general case where $w$ crosses $-1$,
since $w=-1$ corresponds to $f(\rho)=0$, in order that the integral
$\int \d \rho/f(\rho)$ in Eq.~(\ref{EOS4}) is finite,
$f(\rho)$ should behave as
\be
\label{EOS15b}
f(\rho)\sim f_0(\rho - \rho_0)^s\ ,\quad 0<s<1\,,
\ee
where the condition $f(\rho_0)=0$ is assumed.
Since $0<s<1$, $f(\rho)$ should be
multi-valued near $\rho=\rho_0$ in general.

\subsection{Specific model}

In what follows we shall investigate a model characterized by
\be
\label{EOS15}
f(\rho)= \frac{AB \rho^{\alpha+\beta}}{A\rho^\alpha + B \rho^\beta}\,,
\ee
where $A$, $B$, $\alpha$ and $\beta$ are constants.
As is shown below, this dark energy scenario contains a rich
structure from the viewpoint of singularities.

If $\alpha$ is larger than $\beta$, we have
\be
\label{EOS16}
f(\rho)\to \left\{\begin{array}{ll}
A\rho^\alpha \quad & \mbox{when}\ \rho\to 0 \\
B\rho^\beta \quad & \mbox{when}\ \rho\to \infty \\
\end{array} \right. \ .
\ee
When $\alpha,\beta\neq 1$, Eq.~(\ref{EOS4}) gives
\be
\label{EOS17}
a=a_0 \exp \left\{-\frac{1}{3}\left[\frac{\rho^{-\alpha + 1}}{(\alpha - 
1)A}
+ \frac{\rho^{-\beta + 1}}{(\beta - 1)B}\right] \right\}\,.
\ee
For $1>\alpha>\beta$, if $A,B>0$ $(A,B<0)$,
$a$ has a minimum (maximum) $a_0$ at $\rho=0$
and $a$ goes to infinity (vanishes) when $\rho\to \infty$.
For $\alpha>1>\beta$, if $A<0$ and $B>0$ ($A>0$ and $B<0$),
$a$ has a minimum (maximum) at the non-trivial (non-vanishing)
value of $\rho$ and $a$ goes to infinity (zero) when $\rho$ vanishes
or goes to a positive infinity.
For $\alpha>1>\beta$, if $A,B>0$ $(A,B<0)$, $a$ goes to
infinity when $\rho \to \infty$ $(\rho \to 0)$ and
$a$ vanishes when $\rho\to 0$ $(\rho\to \infty)$.
For $\alpha>\beta>1$, $a$ goes to $a_0$ when $\rho\to \infty$.
Furthermore if $A>0$ ($A<0$), $a\to 0$ $(a\to \infty)$ when $\rho\to 0$.
If $A,B>0$ $(A,B<0)$, $a$ is a monotonically
increasing (decreasing) function of $\rho$.
If $A>0$ and $B<0$ ($A<0$ and $B>0$), $a$ has a
nontrivial maximum (minimum) at a finite value of $\rho$.

\subsection{The model (\ref{EOS15}) with $\alpha=2\beta- 1$}

In what follows we shall concentrate on the case that
\be
\alpha=2\beta -1\,.
\ee
Then Eq.~(\ref{EOS17}) may be solved with respect to $\rho$:
\be
\label{EOS18}
\rho =\left\{ - \frac{A}{B}\pm \left(\frac{A^2}{B^2}
- 6A (\beta -1) \ln \frac{a}{a_0}\right)
^{\frac{1}{2}} \right\}^{-\frac{1}{\beta - 1}}\,,
\ee
which is valid for $\beta \neq 1$.
One has $f(\rho)=AB\rho/(A+B)$ for $\beta=1$, which means
that the EOS becomes a usual linear equation
$p=w\rho$ with constant $w$, that is, $w=-1 - AB/(A+B)$.
As this case has been well investigated,
 we will not consider
it in this paper.

When $\alpha=2\beta - 1$, the pressure $p$ is given by
\be
\label{EOS21b}
p=-\rho - \frac{AB \rho^{2\beta -1}}
{A\rho^{\beta -1} + B}\,.
\ee
Now Eq.~(\ref{EOS17}) can be rewritten as
\be
\label{SSS3}
a=a_0 \exp \left\{-\frac{1}{3}\left[
\frac{\rho^{-2(\beta -1)}}{2(\beta - 1)A}
+ \frac{\rho^{-(\beta - 1)}}{(\beta - 1)B}
\right]  \right\}\,.
\ee
Eq.~(\ref{SSS3}) tells that if $\beta>1$, $a\to a_0$ when
$\rho \to \infty$, and $a\to 0$ ($a\to \infty$)
when $\rho\to 0$ and $A>0$ ($A<0$).
On the other hand, if $\beta<1$, $a\to a_0$ when $\rho\to 0$,
and $a\to 0$ ($a\to \infty$) when $\rho\to \infty$ and $A<0$ ($A>0$).
Furthermore Eq.~(\ref{EOS21b}) tells that if $\beta>1$,
\bea
\label{SSS4}
& p\to - \rho - A \rho^{2\beta -1}\ ,\quad & \mbox{when}\quad \rho\to 0\ 
,\\
& p\to - \rho - B \rho^\beta\ ,\quad & \mbox{when}\quad \rho\to \infty\ .
\label{SSS4b}
\eea
Therefore $w=p/\rho \to -1 -0$ ($ -1 +0$) when $\rho\to 0$ and $A>0$ 
($A<0$)
and $w\to +\infty$ ($-\infty$) when $\rho\to \infty$ and $B<0$ ($B>0$).
Thus, except for the case where $\rho\to 0$ and $A<0$,
the dominant energy condition (DEC), which requires,
\be
\label{DEC}
\rho \geq 0\ ,\quad \rho \pm p\geq 0\ ,
\ee
is violated. On the other hand, the strong energy condition (SEC),
\be
\label{SEC}
\rho + 3p\geq 0\ ,\quad \rho + p \geq 0\ ,
\ee
is not violated when $\rho\to \infty$ and $B<0$.
We also note that if $B<0$, the weak energy condition (WEC),
\be
\label{WEC}
\rho\geq 0\ ,\quad \rho + p \geq 0\ ,
\ee
and the null energy condition (NEC),
\be
\label{NEC}
\rho + p \geq 0\ ,
\ee
are not violated, either.

Moreover, if $\beta<1$, one gets
\bea
\label{SSS5}
& p\to - \rho - A \rho^{2\beta -1}\ ,\quad & \mbox{when}\quad \rho\to 
\infty\,, \\
& p\to - \rho - B \rho^\beta\ ,\quad & \mbox{when}\quad \rho\to 0\ .
\label{SSS5b}
\eea
Therefore $w\to +\infty$ ($-\infty$) when $\rho\to 0$ and $B<0$ ($B>0$).
Then DEC is violated but if $B<0$, the violation of SEC does not occur.
When $\rho\to \infty$, one has $w \to -1 -0$ ($ -1+0$)
for $A>0$ ($A<0$).
Then DEC is not violated if $A<0$ although SEC could be violated.
The violation of WEC and NEC does not
occur in this case.

{}From Eq.~(\ref{EOS21b}) it follows that $p$ diverges when
\be
\label{SSS2}
\rho =\rho_s \equiv \left(-\frac{A}{B}\right)^{-\frac{1}{\beta -1}}\,,
\ee
which occurs for $A/B<0$.
Then we have $p\to +\infty$ ($-\infty$) for $B>0$ ($B<0$)
as $\rho \to \rho_s +0$, and $p\to +\infty$ ($-\infty$)
for $B<0$ ($B>0$) as $\rho \to \rho_s -0$.
Then when $\rho\to \rho_s + 0$ with $B>0$ or
$\rho\to \rho_s -0$ with $B<0$,
only the violation of DEC occurs, while other energy
conditions are not violated.

By integrating Eq.~(\ref{tint}), we find
\bea
\label{EOS19}
& & \frac{2}{4\beta - 3} \rho^{-\frac{4\beta - 3}{2}}
+ \frac{2A}{(2\beta - 1)B} \rho^{-\frac{2\beta - 1}{2}} \nn
& &=- \sqrt{3}\kappa A (t-t_0) \equiv \tau\,,
\eea
where $t_0$ is an integration constant.
This is valid for $\beta \neq 1$, $\beta \neq 3/4$, and $\beta \neq 1/2$.
When $\beta=3/4$, instead of Eq.~(\ref{EOS19}), we obtain
\be
\label{EOS19b}
- \ln \frac{\rho}{\rho_0}+
\frac{4A}{B\rho^{\frac{1}{4}}}= \tau \,,
\ee
where $\rho_0$ is a constant introduced for a dimensional reason.
For $\beta=1/2$, it follows
\be
\label{EOS19c}
-2\rho^{\frac{1}{2}} - \frac{A}{B}\ln \frac{\rho}{\rho_0}
= \tau \,.
\ee
We use these relations together with Eqs.~(\ref{EOS18})
and (\ref{EOS21b}) in order to study the properties of singularities.

\subsection{Sudden future singularity for the model (\ref{EOS15}) with  
$A/B<0$}

Let us study the property of the singularity at 
$\rho=\rho_s=(-A/B)^{-1/(\beta- 1)}$.
This corresponds to a finite value of the scale factor by Eq.~(\ref{SSS3}).
When $\rho=\rho_s$, Eqs.~(\ref{EOS19}), (\ref{EOS19b})
and (\ref{EOS19c}) give finite values of $\tau$, i.e.,
\bea
\label{taus1}
\tau_s=-\left(-\frac{A}{B}\right)
^{\frac{4\beta - 3}{2(\beta -1)}}
\frac{4(\beta -1)}{(4\beta - 3)(2\beta -1)}\,,
\eea
for $\beta \neq 3/4, 1/2$, and
\bea
\label{taus2}
\tau_s= -4\left[ {\rm ln}\,\left(-\frac{A}{B\rho_0}\right)+1 \right]\,,
\eea
for $\beta=3/4$, and
\bea
\label{taus3}
\tau_s=\frac{2A}{B}
\left[ 1-{\rm ln}\,\left(-\frac{A}{B\rho_0}\right) \right]\,,
\eea
for $\beta=1/2$.

Since $\rho$ is finite, the Friedmann equation (\ref{FRWH}) also shows that 
$H$
is finite but from Eq.~(\ref{EOS5}),
we find that $\ddot{a}$ diverges because of the divergence of $p$
and the scalar curvature $R$ also diverges since $R=2\kappa^2\left(\rho - 
3p\right)$.
{}From the above arguments the singularity described by Eq.~(\ref{SSS2})
corresponds to the sudden future singularity (type II) in \cite{Barrow}.

\subsection{Classification of singularities for the model (\ref{EOS15})}

In the previous subsection it has been shown that for all values of
$\beta$ the sudden future (type~II) singularity appears
for the model (\ref{EOS15}) if $A/B<0$.
In addition to this there exists a wide variety of singularities depending
on the values of $\beta$.

\subsubsection{$\beta>1$}

When $\beta>1$ Eq.~(\ref{EOS19})  implies that $\rho\to \infty$ as
$\tau\to +0$ ($-0$) for $A/B>0$ ($A/B<0$).
{}From Eqs.~(\ref{SSS3}) and (\ref{SSS4b})
we find $a \to a_0$ and $|p| \to \infty$ as $\rho \to \infty$.
Therefore this corresponds to the type III singularity
at which $\rho$ and $|p|$ diverge with finite $t$ and $a$.
The equation of state behaves as $w \to -1-B\rho^{\beta- 1}$
for $\rho \to \infty$ by Eq.~(\ref{SSS4b}), which means
that $w \to +\infty$ ($-\infty$) for $B<0$ ($B>0$).

Note that $\tau \to \infty$ as $\rho \to 0$
by Eq.~(\ref{EOS19}).
For $A/B>0$ the value of $\tau$ is restricted to be
$0<\tau<+\infty$.
When $A/B<0$, $\tau$ is restricted to be in the range
$\tau > \tau_s$ with negative $\tau_s$
[see Eq.~(\ref{taus1})]. {}From Eq.~(\ref{EOS19})
we find $\d \tau/\d \rho=0$ at $\rho=\rho_s$.
Then two branches appear for the region of $\rho$, i.e.,
$0<\rho<\rho_s$ corresponding to $\tau_s<\tau<+\infty$ and
$\rho_s<\rho<+\infty$ corresponding to $\tau_s<\tau<0$.

\subsubsection{$3/4<\beta<1$}

Let us next consider the case with $3/4<\beta<1$.
We find $\rho\to \infty$ as $\tau \to +0$
from Eq.~(\ref{EOS19}).
Eq.~(\ref{SSS3}) tells that $a\to \infty$ for $A>0$, and
Eq.~(\ref{SSS5}) gives $|p| \to \infty$ for $\rho \to \infty$.
This means that the singularity at $t=t_0$ is the type I
(Big Rip). Since $w \to -1-A\rho^{2(\beta- 1)}$
for $\rho \to \infty$ by Eq.~(\ref{SSS5}), we have
$w \to -1+0$ ($-1-0$) for $A<0$ ($A>0$).

In the limit $\rho \to 0$, one has $\tau \to +\infty$ ($-\infty$)
for $A/B>0$ ($A/B<0$).
The allowed region of $\tau$ is $0<\tau<+\infty$ for $A/B>0$,
whereas $\tau < \tau_s$ with positive $\tau_s$ for $A/B<0$.
In the latter case there exist two branches,
$0<\rho<\rho_s$ corresponding to $-\infty<\tau<\tau_s$ and
$\rho_s<\rho<\infty$ corresponding to $0<\tau<\tau_s$.

\subsubsection{$1/2 \le \beta \le 3/4$}

In the case of $1/2 <\beta < 3/4$, Eq.~(\ref{EOS19}) implies that
$\rho\to 0$ as $\tau\to +\infty$ ($-\infty$) for $A/B>0$ ($A/B<0$).
Meanwhile we have $\tau\to -\infty$ as $\rho\to \infty$.
Then the value of $\tau$ is not limited ($-\infty<\tau<+\infty$)
for $A/B>0$.  We find $-\infty<\tau<\tau_s$ for $A/B<0$,
in which case there exist two branches: $0<\rho<\rho_s$ and
$\rho_s<\rho<\infty$.
Except for the type II singularity at $\tau=\tau_s$ for $A/B<0$,
finite-time singularities do not exist for $1/2 <\beta <3/4$.
This property also holds for $\beta=3/4$ and $\beta=1/2$,
as can be checked by Eqs.~(\ref{taus2}) and (\ref{taus3}).

\subsubsection{$\beta<1/2$}

When $\beta<1/2$, one has $\rho \to \infty$ as $\tau\to -\infty$
and $\rho\to 0$ as $\tau \to -0$ ($+0$) for $A/B>0$ ($A/B<0$).
Then the range of $\tau$ is $-\infty<\tau<0$ for $A/B>0$,
and $0<\tau<\tau_s$ corresponding to $0<\rho<\rho_s$ and 
$-\infty<\tau<\tau_s$
corresponding to $\rho_s<\rho<\infty$ for $A/B<0$.

In the limit $\rho\to 0$ or $\tau\to 0$, Eq.~(\ref{SSS5b}) tells that $p 
\to 0$
for $0<\beta<1/2$, $p \to -B$ for $\beta=0$, and
$p \to +\infty$ ($-\infty$) for $\beta<0$ and $B<0$ ($B>0$).
Therefore there exists a type II singularity for $\beta<0$
($a \to a_0$, $\rho \to 0$, $|p| \to \infty$ as $t \to t_0$).
In this case we have $w \to +\infty$ ($-\infty$) for $B<0$ ($B>0$)
by Eq.~(\ref{SSS5b}).

In the case of $0<\beta<1/2$ one has $\ln a/a_0 \propto \tau^{1- 
\frac{1}{2\beta -1}}$.
Since $1- 1/(2\beta -1)>2$ for $0<\beta<1/2$, $H$ and $\dot{H}$ are finite.
However $\d^n H /\d t^n$ diverge for $n>-1/(2\beta -1)$
as long as $1- 1/(2\beta -1)$ is not an integer.
This corresponds to the type IV singularity in which higher derivatives of
$H$ exhibit divergence even if $a$, $\rho$ and $p$ are finite
as $t \to t_0$.
In this case $w \to +\infty$ ($-\infty$) for $B<0$ ($B>0$)
by Eq.~(\ref{SSS5b}).

When $\beta=0$, we find
$a\sim a_0 \exp \left[B \tau^2/(12 A^2)\right]$
for $\rho \sim 0$,
showing the absence of the
singularity for higher derivatives of $H$.
Therefore there is no any future singularity.
In this case $w \to +\infty$ ($-\infty$) for $B<0$ ($B>0$)
as $\rho \to 0$.

\vspace{1em}

\noindent
The obtained results are summarized as follows:

\noindent
\begin{itemize}
\item  For $A/B<0$ there is always the type II singularity irrespective of 
the
values of $\beta$.
\item  Irrelevant to the sign of $A/B$, the types of singularities
are different depending on the values of $\beta$.

\noindent
\begin{enumerate}

\item $\beta>1$:

There is a type III future singularity. DEC is broken.
$w \to +\infty$ ($-\infty$) for $B<0$ ($B>0$).

\item $3/4<\beta<1$:

There is a type I future singularity for $A>0$.
DEC is broken for $A>0$.
$w \to -1+0$ ($-1-0$) for $A<0$ ($A>0$).

\item $1/2 \le \beta \le 3/4$:

There is no a finite future singularity.

\item $0<\beta<1/2$:

There is a type IV future singularity.
$w \to +\infty$ ($-\infty$) for $B<0$ ($B>0$).

\item $\beta=0$:

There is no finite future singularity,
but when $\rho \to 0$,
$w \to +\infty$ ($-\infty$) for $B<0$ ($B>0$).

\item $\beta<0$:

There is a type II future singularity.
DEC is broken but SEC is not broken for $B<0$.
$w \to +\infty$ ($-\infty$) for $B<0$ ($B>0$).

\end{enumerate}
\end{itemize}

\section{General structure of the singularities}

In this section we discuss the general structure of singularities.
Especially, the relation between the singularities and the behavior of
$f(\rho)$ is clarified.

\subsection{Type I and III singularities}

Let us investigate a situation in which $\rho$ goes to
infinity. Then the Hubble rate diverges from the Friedmann equation
(\ref{FRWH}), which leads to the divergence of all curvatures.
As in Eq.~(\ref{EOS16}), we assume the form
\be
\label{EOS25b}
f(\rho) \to B\rho^\beta\,,~~~{\rm for}~~~
\rho \to \infty\,.
\ee
Here $B$ and $\beta$ are constants.
When $\beta=1$, the EOS is a usual linear equation
$p=w\rho$ with a constant $w (=-1-B)$.

The model with $f(\rho)=B\rho^\beta$ was proposed in Ref.~\cite{final}
and further investigated in Ref.~\cite{Stefancic}.
Since we are now interested in the structure of the
singularities, we assume the behavior (\ref{EOS25b})
with $\beta>0$ only when $\rho$ goes to infinity.
Then by using Eq.~(\ref{EOS4}), it follows
\be
\label{EOS26}
a \sim a_0 \exp \left[\frac{\rho^{1-\beta}}{3(1-\beta)B}
\right]\,.
\ee
When $\beta> 1$, the scale factor remains finite
even if $\rho$ goes to infinity,
which corresponds to the type III singularity \cite{Stefancic}.
When $\beta<1$, we find $a\to \infty$ $(a\to 0)$ as $\rho\to \infty$.
for $B>0$ $(B<0)$.

Since the pressure is now given by
\be
\label{EOS27}
p\sim -\rho - B\rho^\beta\,,
\ee
$p$ always diverges when $\rho$ becomes infinite.
If $\beta>1$, the EOS parameter $w=p/\rho$
also goes to infinity, that is,
$w \to +\infty$ ($-\infty$) for $B<0$ $(B>0$).
When $\beta<1$, we have $w\to -1+0$ ($-1-0$)
for $B<0$ $(B>0$) as $\rho \to \infty$.

By using Eq.~(\ref{tint}) for the function (\ref{EOS25b}), one finds
\be
\label{EOS31}
t\sim t_0 + \frac{2}{\sqrt{3}\kappa B}
\frac{\rho^{-\beta+1/2}}{1-2\beta}\,,~~~
{\rm for}~~~\beta \neq \frac12\,,
\ee
and
\be
\label{EOS32}
t\sim t_0 + \frac{{\rm ln}\,\rho}{\sqrt{3}\kappa B}\,,~~~
{\rm for}~~~\beta=\frac12\,.
\ee
Therefore if $\beta\leq 1/2$, $\rho$ diverges in an infinite future or 
past.
On the other hand, if $\beta>1/2$, the divergence of $\rho$
corresponds to a finite future or past.

{}From the above argument, one can classify the singularities as follows:

\noindent
\begin{enumerate}
\item $\beta> 1$:

There exists a type III singularity.
DEC is broken. $w \to +\infty$ ($-\infty$) if $B<0$ ($B>0$).

\item $1/2<\beta<1$:

There is a type I  future singularity for $B>0$.
DEC is broken in this case.
When $B<0$,
since $a \to 0$ as $\rho \to \infty$,
if the singularity exists in past (future), we may call it Big Bang (Big 
Crunch) singularity.
$w \to -1+0$ ($-1-0$) if $B<0$ ($B>0$).

\item $0< \beta\leq {1}/{2}$:

There is no finite future singularity.
\end{enumerate}

When $\beta<0$, it was shown in Ref.~\cite{Stefancic} that
the type II singularity appears as $\rho \to 0$.
In the next subsection we shall investigate a more general model.

\subsection{Type II singularity}

Lets us consider the type II singularity at which $p$ is singular
but $\rho$ is nonsingular at a finite time $t=t_s$
as in Ref.~\cite{Barrow}.
The starting function $f(\rho)$ is given by
\be
\label{EOS33}
f(\rho)\sim C (\rho_0 - \rho)^{-\gamma}\,,
\ee
where $\gamma$ is a positive constant.
We concentrate on the case where $\rho$ is smaller than $\rho_0$,
but the situation is basically the same for $\rho>\rho_0$
by considering the function $f(\rho)=C (\rho - \rho_0)^{-\gamma}$.
In the limit $\rho \to \rho_0$, the pressure $p$ becomes infinite
because of the divergence of $f(\rho)$.
The scalar curvature $R$ diverges
since $R=2\kappa^2\left(\rho - 3p\right)$.
The EOS is
\be
w\sim -1 - \frac{C}{\rho (\rho_0-\rho)^\gamma}\,,
\ee
which gives $w \to -\infty$ for $C>0$ and
$w \to \infty$ for $C<0$ as $\rho \to \rho_0$.

{}From Eq.~(\ref{EOS4}) the scale factor is integrated to give
\be
\label{EOS34}
a \sim a_0 \exp \left[ - \frac{(\rho_0 - \rho)^{\gamma+1}}
{3C (\gamma + 1)} \right]\,,
\ee
which means that $a$ is finite for $\rho=\rho_0$.
Since the Hubble rate $H$ is nonsingular by Eq.~(\ref{FRWH}),
$\dot{a}$ remains to be finite.
On the other hand Eq.~(\ref{EOS5}) implies that $\ddot{a}$
diverges for $\rho \to \rho_0$.
By using Eq.~(\ref{tint}) we find the following relation
around $\rho \sim \rho_0$:
\be
\label{EOS35}
t \sim t_0 - \frac{(\rho_0-\rho)^{\gamma+1}}
{\kappa C \sqrt{3\rho_0} (\gamma+1)}\,,
\ee
where $t_0$ is an integration constant.
Then $t$ is finite ($t=t_0$) even for $\rho=\rho_0$.
The above discussion shows that the function $f(\rho)$
in Eq.~(\ref{EOS33}) gives rise to the sudden future
singularity.

This type of singularity always appears when the denominator
of $f(\rho)$ vanishes at a finite value of $\rho$.
The model (\ref{EOS25b}) with negative $\beta$ is a special case of
the model (\ref{EOS33}) with $\rho_0=0$.

\section{Attractor solutions}

In this section we account for the contribution of a barotropic
perfect fluid and investigate the attractor properties for
several models presented in Sec.~III.
We study the case of a non-relativistic dark matter
($w_m=0$) without a coupling between
dark matter and dark energy.

\subsection{Constant $p'(\rho)$ with $p'(\rho) \neq -1$}

Let us first investigate the case in which $p'(\rho)$
asymptotically approaches a constant, $p'(\rho) \to w$,
with $w \neq -1$.
One example is presented in Eq.~(\ref{EOS14}), i.e.,
\be
\label{precon}
p=-\rho \pm \frac{2\rho}{3n} \left[
1-\left(\frac{\rho_c}{\rho}\right)^{1/2}
\right]^{1/2}\,,
\ee
where $\rho_c \equiv 48n^2/(\kappa^2 t_s^2)$.
In the limit $\rho \to \infty$ the minus sign in Eq.~(\ref{precon})
gives the constant EOS, $w=-1-2/(3n)<-1$,
whereas the plus sign gives $w=-1+2/(3n)>-1$.

\begin{figure}
\includegraphics[height=3.3in,width=3.5in]{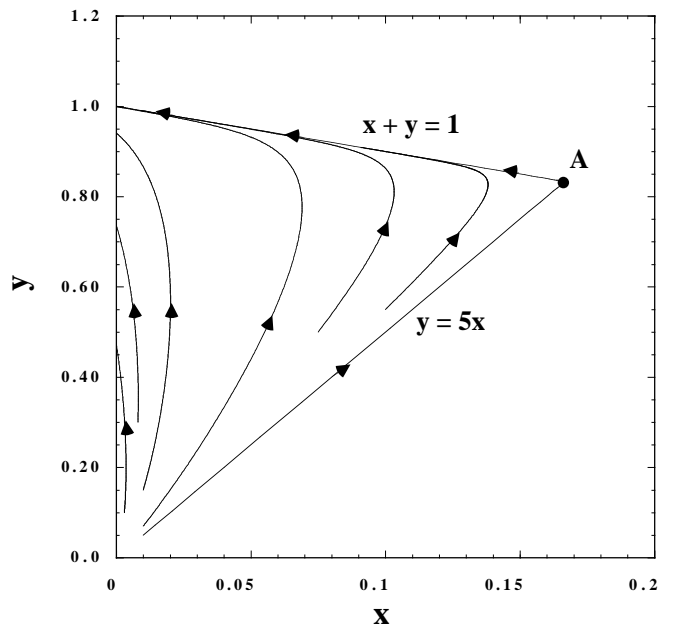}
\caption{The phase plane for the plus sign of the model
(\ref{precon}) with $n=2$. The allowed range corresponds
to $y \le 1-x$, $y>5x$ and $x \ge 0$.
The point A is a saddle
point given by $x=1/(3n)=1/6$ and $y=1-1/(3n)=5/6$.
The solutions approach the region $x=0$, corresponding
to the equation of state: $w=-1$.
}
\label{Fig1}
\end{figure}

When $-1<w<0$ we showed in Sec.~II that the fixed
point $(x, y)=[(1+w)/2, (1-w)/2]$ is a saddle point.
One can easily find that this fixed point is the intersection
of the line (\ref{yandx}) and the line $x+y=1$.
For the model (\ref{precon}) with a plus sign,
one has $y/x+1=3n/\sqrt{1-(\rho_c/\rho)^{1/2}}$ by Eq.~(\ref{weq}).
Since $\sqrt{1-(\rho_c/\rho)^{1/2}}<1$, the range of $x$
and $y$ in the phase plane is restricted to be $y > (3n-1)x$,
$y \le 1-x$ and $x \ge 0$.
Note that the line (\ref{yandx}) corresponds to
$y=(3n-1)x$ by setting $w=-1+2/(3n)$.
Therefore the intersection of two lines $y=(3n-1)x$ and
$y=1-x$ is a saddle point given by $(x, y)=[1/(3n), 1-1/(3n)]$.
This is plotted as a point A in Fig.~\ref{Fig1}.

We run our numerical code for initial conditions of $x$ and
$y$ that belong to the allowed range explained above.
We find that the solutions gradually approach $x=0$,
which gives the EOS: $w=-1$ by Eq.~(\ref{weq}).
The point A corresponding to
$w=-1+2/(3n)$ is not an attractor point.
As the trajectories approach $x=0$, the energy density $\rho$
decreases to a minimum value $\rho_c$.
Figure \ref{Fig1} shows that there is some portion of the
energy density of a barotropic fluid ($\Omega_m=1-x-y>0$)
around $x=0$ if the initial condition of $x$ is much smaller
than $y$. When the trajectories get closer to the line $x+y=1$ in the
middle of the way, the final attractor point is $(x, y)=(0, 1)$,
corresponding to a universe described by a cosmological constant
with a vanishing energy density of a barotropic fluid.

\begin{figure}
\includegraphics[height=3.3in,width=3.5in]{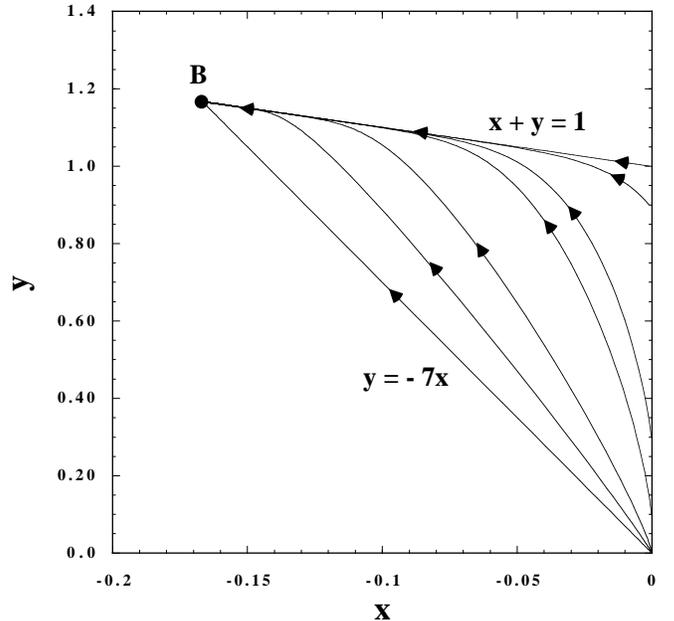}
\caption{The phase plane for the minus sign of the model
(\ref{precon}) with $n=2$.
The allowed range corresponds
to $y \le 1-x$, $y>-7x$ and $x \le 0$.
The point B is a stable node given by $x=-1/(3n)=-1/6$ and
$y=1+1/(3n)=7/6$, which corresponds to the late-time
attractor. The equation of state at this point is a constant:
$w=-1-2/(3n)=-4/3$.
 }
\label{Fig2}
\end{figure}

When $w<-1$ there exists one stable fixed
point: $(x, y)=[(1+w)/2, (1-w)/2]$.
In scalar-field dark energy models, the kinematic term is
negative ($x<0$) at this critical point.
For the model (\ref{precon}) with a minus sign,
one can easily show that the allowed range of $x$ and $y$
is described by $y > -(3n+1)x$, $y \le 1-x$ and
$x \le 0$. The intersection of the two lines $y=-(3n+1)x$
and $y=1-x$, i.e., $(x, y)=[-1/(3n), 1+1/(3n)]$,
represents the above stable fixed point.
In Fig.~\ref{Fig2} we plot the phase plane in terms of
$x$ and $y$ for the model (\ref{precon}) with a minus sign.
The solutions approach the above stable fixed point at which
the EOS is given by $w=-1-2/(3n)<-1$.
We recall that this attractor point corresponds to
the Big Rip singularity with a divergent energy density
at finite time $t_s$.
If we connect two functions (\ref{precon}) at the minimum
value of $\rho$, the solutions starting from the region
$x > 0$ approach the line $x=0$, enter the region
$x<0$, and then approach the above stable fixed point.
It is interesting to note that the final attractor point is the
Big Rip type even if the system begins from a non-phantom
stage ($x>0$).

\subsection{$|p'(\rho)| \to \infty$}

Let us next consider the case in which $|p'(\rho)|$
diverges asymptotically. One example is given by
the model (\ref{EOS21b}) with $\beta>1$.
In this case there exists a type III singularity at
which $\rho$ and $p$ exhibit divergence with
finite time and scale factor.
In the limit $\rho \to \infty$ the pressure $p$
is given by Eq.~(\ref{SSS4b}), which means that
$p'(\rho) \sim -1-\beta B \rho^{\beta-1} \to -\infty$
and $w \sim -1-B\rho^{\beta-1} \to -\infty$
for $B>0$.

We shall study a situation in which the system starts out from
the region $x<0$ (i.e., negative $w$) and then evolves toward
$w \to -\infty$. By Eqs.~(\ref{weq}) and (\ref{EOS21b})
one obtains $y/x=-1-2(Au+B)/(ABu^2)$, where
$u \equiv \rho^{\beta- 1}>0$. When $A$ and $B$ are
both positive, we find $y>-x$ for $x<0$.
Therefore the allowed range is restricted to be
$y>-x$, $y \le 1-x$ and $x<0$ in the phase plane.

\begin{figure}
\includegraphics[height=3.3in,width=3.5in]{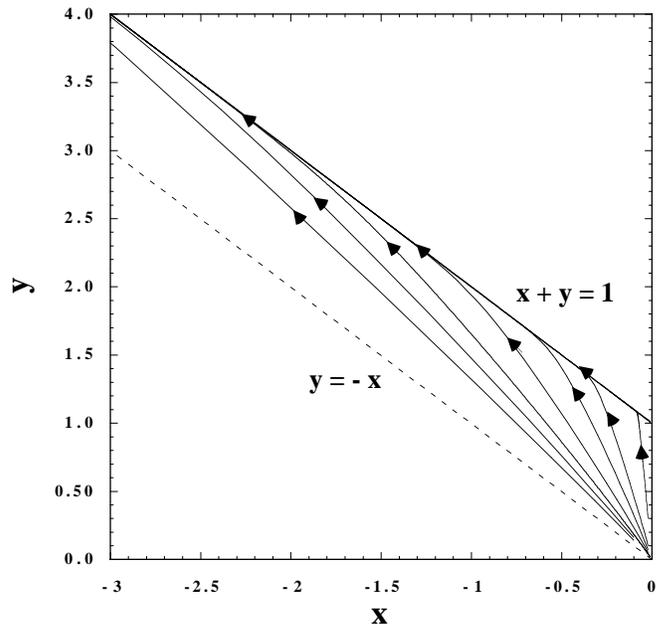}
\caption{The phase plane for the model (\ref{EOS15}) with
$\beta=1.1$, $\alpha=2\beta- 1$, $A=2$ and $B=1$.
The solutions approach ``instantaneous'' critical points:
$x=(1+p'(\rho))/2$ and $y=(1-p'(\rho))/2$, which diverge
as $x \to -\infty$ and $y \to \infty$ at the type III singularity.
}
\label{Fig3}
\end{figure}

In Fig.~\ref{Fig3} we plot the trajectories of the solutions
for the model (\ref{EOS21b}) with $\beta=1.1$, $A=2$
and $B=1$. We find that all solutions approach
$x \to -\infty$ and $y \to +\infty$ along
with the line $x+y=1$.
The points $(x, y)=[(1+p'(\rho))/2, (1-p'(\rho))/2]$ can be
regarded as ``instantaneous'' critical points corresponding
to the evolving critical point (ii) introduced in Sec.~II.
In fact these points are on the line $x+y=1$ and all
trajectories approach this line as illustrated in Fig.~\ref{Fig3}.
We note that along the line $x+y=1$ the energy density of
the barotropic fluid is vanishing ($\Omega_m=0$).

\subsection{$p'(\rho) \to -1$}

Finally we consider the situation in which $p'(\rho)$
approaches $-1$ asymptotically. One example is provided
by the model (\ref{EOS21b}) with $3/4<\beta<1$,
in which case there is a type I singularity.
Since $p$ is given by Eq.~(\ref{SSS5}) in the limit
$\rho \to \infty$, we have $p'(\rho) \sim -1-
A (2\beta- 1) \rho^{2(\beta -1)} \to -1-0$
and $w \sim -1-A\rho^{2(\beta-1)} \to -1-0$ for $A>0$.

\begin{figure}
\includegraphics[height=3.3in,width=3.5in]{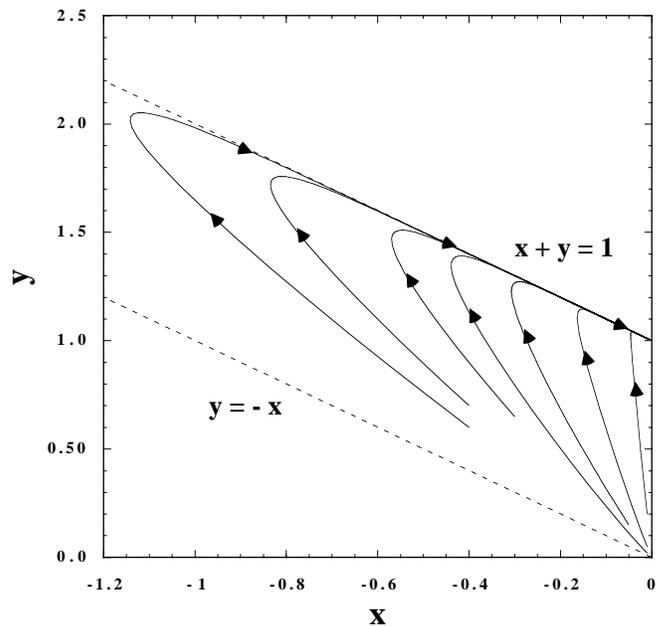}
\caption{The phase plane for the model (\ref{EOS15}) with
$\beta=0.85$, $\alpha=2\beta- 1$, $A=2$ and $B=1$.
The solutions approach the type I singularity at which
the equation of state is $w=-1$ with $x=0$ and $y =1$. }
\label{Fig4}
\end{figure}

We study the case in which both $A$ and $B$ are positive.
When $x<0$ the allowed range of $x$ and $y$ is the same as
in the case of the subsection B.
We show in Fig.~\ref{Fig4} the trajectories of the solutions
for $\beta=0.85$, $A=2$ and $B=1$.
The solutions eventually approach the point $(x, y)=(0, 1)$
along with the line $x+y=1$.
This indicates that the trajectories evolve on instantaneous
critical points $(x, y)=[(1+p'(\rho))/2, (1-p'(\rho))/2]$
with the increase of $p'(\rho)$ toward $p'(\rho)=-1$.
The final state of the universe is dominated by a cosmological
constant with vanishing $\Omega_m$.

\section{Coupled phantom scenarios}

In this section we shall investigate a situation in which
dark energy is coupled to dark matter.
The coupled quintessence scenario was originally proposed
in Ref.~\cite{Luca} as an extension of nonminimal coupling
theories \cite{Luca2}.
This scenario has an advantage to explain the coincidence problem,
that is, why the energy density of dark energy is
comparable with the energy density of dark matter or usual matter.
In Ref.~\cite{Luca} the coupling between a scalar field $\phi$
responsible for dark energy and ordinary matter is
assumed to take the forms:
$T^{\mu}_{\nu (\phi); \mu}=CT_{(m)}\phi_{;\nu}$ and
 $T^{\mu}_{\nu (m); \mu}=-CT_{(m)}\phi_{;\nu}$, where
$T_{\mu \nu (\phi)}$ and $T_{\mu \nu (m)}$ are
the energy momentum tensors of $\phi$ and ordinary matter,
respectively.
Recently, a system of a phantom scalar field coupled to
dark matter was studied in Ref.~\cite{Cai} by writing the
FLRW equations in an autonomous form,
but this is not equivalent to our analysis since we assumed
the EOS of dark energy rather than the existence of a scalar field.
Furthermore the forms of the coupling $Q$ are different from
the ones we consider in this paper.

\subsection{Analytic solutions for coupled phantom}

Let the dark energy be a fluid whose equation of motion
is given by Eq.~(\ref{rhoeq}).
In general the coupling $Q$ can be the function of  $a$, $H$, $\rho_m$,
$\dot{\rho}_m$, $\rho$ and $\dot{\rho}$.
We can express the coupling $Q$ that corresponds to scaling solutions,
that is, the ratio $r=\rho_m/\rho$ is a constant.
By Eqs.~(\ref{rhoeq}) and (\ref{rhoeq2}) we find
\be
\dot{r}=r \left[\frac{Q}{\rho_m}+\frac{Q}{\rho}
-3H(w_m-w)\right]\,.
\ee
We obtain the coupling $Q$ for the existence of scaling
solutions \cite{ZP,GZ}, as
\be
Q=3H(w_m-w) \frac{\rho \rho_m}{\rho +\rho_m}\,,
\ee
which is derived by setting $\dot{r}=0$.
In this case the energy density of ordinary matter does not vanish
asymptotically.

Several authors considered the coupling of the forms like
$Q=3H \delta (\rho +\rho_m)$ \cite{GZ} and
$Q=\delta H \rho_m$ \cite{CW}, where $\delta$ is a coupling
strength. In this work we are interested in the coupling with which
singularities discussed in preceding sections can be obtained
analytically.

Let us consider the following coupling as a toy model:
\be
\label{EOS37}
Q=\delta H^2 \,,
\ee
where $\delta$ is a constant.
As usually dark matter is regarded as a dust, i.e.,
\be
\label{EOS38}
p_m=0\,.
\ee
The following EOS for dark energy is assumed:
\be
\label{EOS39}
p=-2\rho \,,
\ee
that is, $f(\rho)=\rho$ in Eq.~(\ref{EOS1}).

Then combining Eq.~(\ref{FRWH}) with Eqs.~(\ref{rhoeq}) and
(\ref{rhoeq2}), we find a solution
\bea
\label{EOS40}
H &=& \frac{2}{3}\left(\frac{1}{t} + \frac{1}{t_s - t}\right)\ , \\
\label{EOS40a2}
\rho_m &=& \frac{4}{3\kappa^2}\left(\frac{1}{t} + \frac{1}{t_s - 
t}\right)\frac{1}{t}\ , \\
\label{EOS40a}
\rho &=& \frac{4}{3\kappa^2}\left(\frac{1}{t} + \frac{1}{t_s - t}\right) 
\frac{1}{t_s - t} \,,
\eea
where
\be
\label{EOS40b}
t_s\equiv \frac{9}{\delta \kappa^2}\,.
\ee
The same solution can be obtained if, instead of Eq.~(\ref{EOS37}),
 the following coupling is adopted
\be
\label{EOS41}
Q=\frac{9H\rho \rho_m}{2\left(\rho + \rho_m\right)}\ .
\ee
In this case $t_s$ appears as an undetermined constant.
The solution (\ref{EOS40}) is the same as Eq.~(\ref{EOS8})
with the choice $n=2/3$.
Then the above solution gives rise to the Big Rip singularity at $t=t_s$.

In the absence of the coupling $Q$, it is rather difficult to realize a 
model
in which the equation of state $w$ changes from $w>-1$ to $w<-1$
unless the function $f(\rho)$ is double-valued as discussed in Sec.~III.
However this can be realized if the phantom is coupled to
dark matter with an appropriate coupling.
We note that the ratio $r=\rho_m/\rho$ behaves as $r=(t_s-t)/t$
by Eqs.~(\ref{EOS40a2}) and (\ref{EOS40a}), which means
that this ratio is dynamically changing and vanishes at $t=t_s$.
Therefore the solution (\ref{EOS40}) does not correspond to scaling
solutions. This is understandable, since the system is completely
dominated by the energy density of dark energy around the Big Rip.

\subsection{Multiple scalar field model}

It is possible to construct a scalar-field model which reproduces
the solutions (\ref{EOS40})-(\ref{EOS40a}).
This can be realized by a four-dimensional Lagrangian
with two scalar fields $\phi$ and $\chi$:
\bea
\label{EOS49}
{\cal L} = \frac{1}{2\kappa^2}R - \frac{1}{2}(\nabla \phi)^2
+ \frac{1}{2}(\nabla \chi)^2-V(\phi, \chi)\,,
\eea
where the effective potential is
\bea
\label{effpo}
\hspace*{-0.5em}
V(\phi, \chi)=\frac{\eta^2}{t_0^2} \e^{-\frac{\phi + \chi}{\phi_0}}
+\frac{\eta^2 - \phi_0^2}{2t_0^2} \e^{-\frac{2\phi}{\phi_0}}
+\frac{\eta^2 + \phi_0^2}{2t_0^2} \e^{-\frac{2\chi}{\phi_0}}.
\eea
Here $\phi_0$ and $t_0$ are constants, and
\bea
\label{etadef}
\eta^2=\frac32 \kappa^2 \phi_0^4\,.
\eea

In Eq.~(\ref{EOS49}) the sign of the kinetic term of $\chi$ is opposite
compared to the canonical one, which tells that the field $\chi$ is
regarded as a phantom.
Then in this model, if we consider the quantization, the problem of 
negative
 norm states
would occur as the phantom is ghost \cite{CHT,Cline,Piazza}.
When we adopt the Brans-Dicke type theory as in Ref.~\cite{ENO},
it is possible to avoid the problem of the negative sign in the kinetic 
term.
Since it makes, however, all things
complicated, we pay our attention to the scenario given by  (\ref{EOS49})
in order to show how the solutions (\ref{EOS40})-(\ref{EOS40a})
are realized in a simple scalar field model.

The total energy density $\rho_t$ corresponding to the
Lagrangian (\ref{EOS49}) is given by
\bea
\label{EOS50}
\rho_t &=& \frac{1}{2}\dot{\phi}^2-
\frac{1}{2}\dot{\chi}^2
 + \frac{\eta^2}{t_0^2} \e^{-\frac{\phi + \chi}{\phi_0}} \nn
&& + \frac{\eta^2 - \phi_0^2}{2t_0^2} \e^{-\frac{2\phi}{\phi_0}}
 + \frac{\eta^2 + \phi_0^2}{2t_0^2}  \e^{-\frac{2\chi}{\phi_0}}\,.
\eea
In the FRW background the
equations of motion for scalar fields are
\bea
\label{EOS51}
& & \ddot{\phi}+3H\dot{\phi}-\frac{\eta^2}{\phi_0 t_0^2}
\e^{-\frac{\phi + \chi}{\phi_0}}-
\frac{\eta^2 - \phi_0^2}{\phi_0 t_0^2}
\e^{-\frac{2\phi}{\phi_0}}=0\,, \\
& & \ddot{\chi}+3H\dot{\chi}+\frac{\eta^2}{\phi_0 t_0^2}
\e^{-\frac{\phi + \chi}{\phi_0}}+
\frac{\eta^2+\phi_0^2}{\phi_0 t_0^2}
\e^{-\frac{2\chi}{\phi_0}}=0\,.
\eea
Then by using the Friedmann equation $H^2=\kappa^2 \rho_t/3$, 
we find a solution:
\bea
\label{EOS52}
& & H=\frac{\kappa^2\phi_0^2}{2}\left(\frac{1}{t} + \frac{1}{t_s - 
t}\right)\,, \\
& & \phi=\phi_0 \ln \frac{t}{t_0}\,,~~~\chi = \phi_0 \ln \frac{t_s - 
t}{t_0}\,,
\eea
where $t_s$ appears as a constant of the integration.
Then if one chooses $\phi_0$ as
\be
\label{EOS48}
\phi_0^2=\frac{4}{3\kappa^2}\ ,
\ee
the Hubble rate $H$ in Eq.~(\ref{EOS40}) is reproduced.

The field $\phi$ is identified as dark matter with an energy density
$\rho_m$ and the field $\chi$ as dark energy with an energy density
$\rho$. The total energy density $\rho_t$ in Eq.~(\ref{EOS50})
may be given by
\bea
\label{EOS53}
\rho_m &=& \frac{1}{2} \dot{\phi}^2
+ \frac{\eta^2}{2t_0^2} \e^{-\frac{\phi+\chi}{\phi_0}}
+ \frac{\eta^2 - \phi_0^2}{2t_0^2} \e^{-\frac{2\phi}{\phi_0}} \nn
&=& \frac{\eta^2}{2}\left(\frac{1}{t} + \frac{1}{t_s - t}\right)
\frac{1}{t}\,, \nn
\rho &=& - \frac{1}{2} \dot{\chi}^2
+ \frac{\eta^2}{2t_0^2} \e^{-\frac{\phi + \chi}{\phi_0}}
+ \frac{\eta^2 + \phi_0^2}{2t_0^2}  \e^{-\frac{2\chi}{\phi_0}} \nn
&=& \frac{\eta^2}{2}\left(\frac{1}{t} + \frac{1}{t_s - t}\right)
\frac{1}{t_s-t}\,,
\eea
which agree with Eqs.~(\ref{EOS40a2}) and (\ref{EOS40a})
by using Eqs.~(\ref{etadef}) and (\ref{EOS48}).
Furthermore if we define the pressures $p_m$ and $p$ as
\bea
\label{EOS54}
p_m &=& \frac{1}{2}\dot{\phi}^2
 - \frac{\eta^2}{2t_0^2} \e^{-\frac{\phi + \chi}{\phi_0}}
 - \frac{\eta^2 - \phi_0^2}{2t_0^2} \e^{-\frac{2\phi}{\phi_0}} \ , \nn
p &=& - \frac{1}{2}\dot{\chi}^2
 - \frac{\eta^2}{2t_0^2} \e^{-\frac{\phi + \chi}{\phi_0}}
 - \frac{\eta^2 + \phi_0^2}{2t_0^2}  \e^{-\frac{2\chi}{\phi_0}},
\eea
it follows
\bea
\label{EOS55}
&& \dot{\rho}_m + 3H\left(\rho_m + p_m\right)
= - \left\{ \dot{\rho} + 3H\left(\rho + p\right) \right\} \nn
&& = \frac{\eta^2}{2\phi_0 t_0^2}\e^{-\frac{\phi + \chi}{\phi_0}}
\left(\dot{\phi}-\dot{\chi} \right)\,.
\eea
By comparing Eq.~(\ref{EOS55}) with
Eqs.~(\ref{rhoeq}) and (\ref{rhoeq2}), one gets
\be
\label{EOS56}
Q=\frac{\eta^2}{2\phi_0 t_0^2}\e^{-\frac{\phi + \chi}{\phi_0}}
\left(\dot{\phi}-\dot{\chi} \right)\,.
\ee

Note that all the parameters in the action (\ref{EOS49}) have
the scale of order the Planck mass.
The parameter $t_s$ appears dynamically as a constant of the integration.
If the present universe corresponds to the time $t$ of order $t_s$,
one can estimate $t_s$ as
$t_s \sim 1.8\times 10^{10}\,\mbox{years} \sim \left(10^{-33}
\mbox{eV}\right)^{-1}$.
The coupling $Q$ in Eq.~(\ref{EOS56}) cannot be straightly rewritten
in the form of Eq.~(\ref{EOS37}) or Eq.~(\ref{EOS41}),
but with the solution, all the expressions for $H$, $\rho_m$ and
$\rho$ give the same functions in terms of the time $t$.

One can consider a generalization of the model in Eq.~(\ref{EOS49})
as follows:
\bea
{\cal L}&=& \frac{1}{2\kappa^2}R - \frac{1}{2} (\nabla \phi)^2
+ \frac{1}{2} (\nabla \chi)^2-\frac{\eta\theta}{t_0^2}
\e^{-\frac{\phi}{\phi_0} - \frac{\chi}{\zeta_0}} \nn
& & -\frac{\eta^2 - \phi_0^2}{2t_0^2} \e^{-\frac{2\phi}{\phi_0}}
 - \frac{\theta^2 + \chi_0^2}{2t_0^2} \e^{-\frac{2\chi}{\chi_0}}\,.
\eea
Here $\phi_0$, $\chi_0$, and $t_0$ are constants and
\be
\label{EOS59}
\eta^2 = \frac32 \kappa^2\phi_0^4\ , \quad
\theta^2 = \frac32\kappa^2\chi_0^4\,.
\ee
Then a solution has the following form:
\bea
\label{EOS60}
&& H=\frac{\kappa^2}{2}\left(\frac{\phi_0^2}{t} + \frac{\chi_0^2}{t_s - 
t}\right)\,, \\
&& \phi=\phi_0 \ln \frac{t}{t_0}\,,~~~\chi = \chi_0 \ln \frac{t_s - 
t}{t_0}\,.
\eea
The Hubble parameter has a minimum at $t=\phi_0 t_s/(\phi_0 +\chi_0)$
for $\chi_0/\phi_0>0$ and diverges for $t \to 0$ and $t \to t_s$.
When $t \ll t_s$  the system is dominated by the field $\phi$ and
the effective EOS is given by
$w \sim -1+4/(3\kappa^2 \phi_0^2)$.
On the other hand the field $\chi$ dominates around $t \sim t_s$,
corresponding to the equation of state:
$w \sim -1-4/(3\kappa^2 \chi_0^2)$.
Then the system transits from the region $w>-1$ to the region $w<-1$
and approaches the Big Rip singularity at $t=t_s$.

We may investigate whether the present model can satisfy the present
observational data.
The deceleration of the universe changed to the acceleration about
five billion years ago.
Let us write the corresponding time as $t=t_c$.
Since
\bea
\label{3S1}
\ddot a/a &=&H^2 + \dot H \nn
&=& \frac{\kappa^2}{4}\left\{\frac{\left(\kappa^2 \phi_0^2 - 
2\right)\phi_0^2}{t^2}
+ \frac{\left(\kappa^2 \chi_0^2 + 2\right)\chi_0^2}{\left(t_s - t\right)^2} 
\right. \nn
&& \left. + \frac{2\kappa^2\phi_0^2 \chi_0^2}{t\left(t_s - 
t\right)}\right\}\,,
\eea
then $t_c$ satisfies
\bea
\label{3S2}
&& \left(\phi_0^2 - \chi_0^2\right)\left\{\kappa^2
\left(\phi_0^2 - \chi_0^2\right) -2 \right\}t_c^2 \\
&& - 2\phi_0^2 \left\{\kappa^2\left(\phi_0^2 - \chi_0^2\right) -2 
\right\}t_s t_c
+ \left(\kappa^2 \phi_0^2 - 2\right)\phi_0^2 t_s^2=0\,. \nonumber
\eea
According to the present data, the density of dark energy is 72$\%$ of
the total density and that of dark matter is 21$\%$ \cite{WMAP}.
We define the energy density of dark matter $\rho_m$ and that of
the phantom $\rho$ by
\bea
\label{3S3}
\rho_m &=& \frac{1}{2}{\dot \phi}^2 + 
\frac{\eta\theta}{2t_0^2}\e^{-\frac{\phi}{\phi_0}
- \frac{\chi}{\chi_0}}
+ \frac{\eta^2 - \phi_0^2}{2t_0^2}\e^{-2\frac{\phi}{\phi_0}} \nn
&=& \frac{3\kappa^2\phi_0^2}{4t}\left(\frac{\phi_0^2}{t}
+ \frac{\chi_0^2}{t_s - t}\right)\,,\nn
\rho &=& - \frac{1}{2}{\dot \chi}^2 + 
\frac{\eta\theta}{2t_0^2}\e^{-\frac{\phi}{\phi_0}
- \frac{\chi}{\chi_0}}+ \frac{\theta^2 +
\chi_0^2}{2t_0^2}\e^{-2\frac{\chi}{\chi_0}} \nn
&=& \frac{3\kappa^2\chi_0^2}{4\left(t_s - t\right)}\left(\frac{\phi_0^2}{t}
+ \frac{\chi_0^2}{t_s - t}\right)\,.
\eea

If $t$ corresponds to the present age of the universe,
that is, fourteen billion years, one finds
\be
\label{EOS42}
\frac{t\chi_0^2}{\left(t_s - t\right)\phi_0^2}\sim \frac{7}{2}\,.
\ee
Then
\bea
\label{EOS43}
t&\sim&\left(\frac{2\chi_0^2}{7\phi_0^2} + 1\right)^{-1}t_s\, ,\\
t_c&\sim&\frac{14 - 5}{14}t\sim \frac{9}{14}
\left(\frac{2\chi_0^2}{7\phi_0^2} + 1\right)^{-1}t_s\,.
\label{EOS43b}
\eea
If we define a coincidence time $\hat t$
by $\rho=\rho_m$, one has
\be
\label{Coin}
\hat t = \left( 1 + \frac{\chi_0^2}{\phi_0^2}\right)\,.
\ee

Dark matter is regarded as a nonrelativistic matter with EOS: $w_m=0$.
We may consider a situation in which this originates from the field
$\phi$ with constant EOS: $w_m=-1+4/(3\kappa^2 \phi_0^2)$,
which gives $\phi_0^2=4/(3\kappa^2)$.
We caution that this property does not necessarily hold, since
this constant EOS does not correspond to the attractor
solution as shown in Sec.~V.
We just wish to estimate the approximate time
$t_s$ by imposing this condition.
Then substituting $\phi_0^2=4/(3\kappa^2)$ and Eq.~(\ref{EOS43b})
for Eq.~(\ref{3S2}), we obtain
\be
\label{3S4}
\kappa^2 \chi_0^2= 0.0831097...
\ee
This gives $t_s - t\sim 0.02 t \sim 0.3$ billion years toward the Big Rip,
which is rather a short time.
The coincidence time $\hat t$ is given
by $t-\hat t \sim 0.04 t \sim 0.6$
billion years ago.

\section{Account of quantum effects around the singularities}

It is quite natural to investigate the role of quantum effects near
singularities. Indeed, the dark energy universe with a singularity
typically evolves to its end with the growth of the energy.
As a consequence the curvature of the universe grows as well,
which implies the beginning of a second quantum gravity era.
The type IV singularity introduced in Sec.~III is not of this sort
because the curvature does not grow significantly there.

In what follows we shall study the effect of quantum backreaction
of conformal matter around the type I (Big Rip), type II (sudden future)
and type III singularities.
In these cases, the curvature of the universe becomes large
around the singularity at $t=t_s$,
although the scale factor $a$ is finite for type II and III singularities.
Since quantum corrections usually contain the powers of the curvature
or higher derivative terms, such correction terms play important roles
near the singularity. We now include the quantum effects by taking into
account the contribution of conformal anomaly
as a backreaction.
The conformal anomaly $T_A$ has the following form:
\be
\label{OVII}
T_A=b\left(F+{2 \over 3}\Box R\right) + b' G + b''\Box R\ ,
\ee
where $F$ is the square of a 4d Weyl tensor,
$G$ is a Gauss-Bonnet curvature invariant, which are given as
\bea
\label{GF}
F&=& (1/3)R^2 -2 R_{ij}R^{ij}+ R_{ijkl}R^{ijkl}\,, \nn
G&=&R^2 -4 R_{ij}R^{ij}+ R_{ijkl}R^{ijkl}\,.
\eea
In general, with $N$ scalar, $N_{1/2}$ spinor, $N_1$ vector fields, $N_2$ 
($=0$ or $1$)
gravitons and $N_{\rm HD}$ higher derivative conformal scalars,
the coefficients $b$ and $b'$ are given by
\bea
\label{bs}
\hspace*{-2.2em} &&
b={N +6N_{1/2}+12N_1 + 611 N_2 - 8N_{\rm HD} \over 120(4\pi)^2}\,, \nn
\hspace*{-2.2em} &&
b'=-{N+11N_{1/2}+62N_1 + 1411 N_2 -28 N_{\rm HD} \over 360(4\pi)^2}\,.
\eea
We have $b>0$ and $b'<0$ for the usual matter except for
higher derivative conformal scalars.
Notice that $b''$ can be shifted by a finite renormalization of the
local counterterm $R^2$, so  $b''$ can be arbitrary.

In terms of the corresponding energy density $\rho_A$ and the
pressure density $p_A$,
$T_A$ is given by $T_A=-\rho_A + 3p_A$.
Using the energy conservation law in FLRW universe:
\be
\label{CB1}
\dot{\rho}_A+3 H\left(\rho_A + p_A\right)=0\,,
\ee
we may delete $p_A$ as
\be
\label{CB2}
T_A=-4\rho_A -\dot{\rho}_A/H \,.
\ee
This gives the following expression for $\rho_A$:
\bea
\label{CB3}
\hspace*{-0.4em} \rho_A&=& -\frac{1}{a^4} \int \d t\, a^4 H T_A \nn
\hspace*{-0.4em}&=&  -\frac{1}{a^4} \int \d t\, a^4 H \Bigl[-12b \dot{H}^2
+ 24b' (-\dot{H}^2 + H^2 \dot{H} + H^4)  \nn
\hspace*{-0.5em}& &- (4b + 6b'')\left(\dddot{H}
+ 7 H \ddot{H} + 4\dot{H}^2 +
12 H^2 \dot{H} \right) \Bigr]\,.
\eea

A different form of $\rho_A$ was obtained in Ref.~\cite{NOqc}
by requiring that the quantum corrected energy momentum tensor
$T_{A\,\mu\nu}$ has the form as $T_{A\,\mu\nu}=(T_A/4)g_{\mu\nu}$
in the conformal metric case rather than
assuming the conservation law (\ref{CB1}).
The quantum corrected FLRW equation is
\be
\label{EOSq1}
\frac{3}{\kappa^2}H^2=\rho + \rho_A\,.
\ee
Since the curvature is large around $t=t_s$,
one may assume $(3/\kappa^2)H^2\ll \left| \rho_A\right|$.
Then  $\rho\sim - \rho_A$ from Eq.~(\ref{CB3}),
which gives
\bea
\label{EOSq2}
\hspace*{-2.2em}&& \dot{\rho}+4H\rho \nn
\hspace*{-2.2em}&& =  H\Bigl[-12b \dot{H}^2
+ 24b' (-\dot{H}^2 + H^2 \dot{H} + H^4)  \nn
\hspace*{-2.2em}& &- (4b + 6b'')\left(\dddot{H}
+ 7 H \ddot{H} + 4\dot{H}^2 +
12 H^2 \dot{H} \right) \Bigr]\,.
\eea
{}From the energy conservation law $\dot{\rho}+
3H\left(\rho + p\right)=0$ for $p=-\rho - f(\rho)$,
one obtains
\be
\label{EOSq3}
H=\frac{\dot{\rho}}{3 f(\rho)}\,.
\ee

\subsection{Type III singularity}

For the type III singularity, the energy density $\rho$ diverges at 
$t=t_s$.
Let us consider the function $f(\rho)\sim B\rho^\beta$ discussed
in Sec.~IV. In the absence of quantum corrections
the type III singularity appears for $\beta>1$ and $\rho$ behaves
as $\rho \propto \left(t_s - t\right)^{\frac{2}{1-2\beta}}$.
When quantum corrections are taken into account,
it is natural to assume that near the singularity $\rho$ behaves as
\be
\label{EOSq4}
\rho = \rho_0\left(t_s - t\right)^{\tilde\gamma}\,.
\ee
As $\rho$ may diverge at $t=t_s$, we consider negative values of
$\tilde\gamma$.
Using Eq.~(\ref{EOSq3}), one finds
\be
\label{EOSq4b}
H =- \frac{\tilde\gamma \rho_0^{1-\beta}}{3B}
\left(t_s - t\right)^{-1 + \tilde\gamma\left(1 - \beta\right)}\,.
\ee

Since we are considering the case $\beta>1$ and $\tilde\gamma<0$,
we have that $\tilde\gamma\left(1 - \beta\right)>0$.
By picking up most singular term in the r.h.s of Eq.~(\ref{EOSq2}), it 
follows
\be
\label{EOSq6}
\dot{\rho} \sim -6\left(\frac{2}{3}b + b''\right)H \dddot{H}\,.
\ee
Then substituting Eqs.~(\ref{EOSq4}) and (\ref{EOSq4b}) for
Eq.~(\ref{EOSq6}), we obtain
\be
\label{gam}
\tilde\gamma=\frac{4}{1-2\beta}\,.
\ee
This means that $\rho$ and $H$ evolve as
\be
\label{rhoHsolu}
\rho \propto (t_s-t)^{\frac{4}{1-2\beta}}\,,~~~~
H \propto (t_s-t)^{\frac{3-2\beta}{1-2\beta}}\,,
\ee
around $t=t_s$.
Compared with the classical evolution:
$\rho \propto \left(t_s - t\right)^{\frac{2}{1-2\beta}}$,
the energy density diverges more rapidly because of
the condition: $4/(1-2\beta)<2/(1-2\beta)<0$ for $\beta>1$.
On the other hand the Hubble rate behaves as
$H \propto (t_s-t)^{\frac{1}{1-2\beta}}$ in the classical case,
which means that $H$ is less singular in the presence of
quantum corrections because of the condition:
$(3-2\beta)/(1-2\beta)>1/(1-2\beta)$ for $\beta>1$.

\begin{figure}
\includegraphics[height=3.3in,width=3.5in]{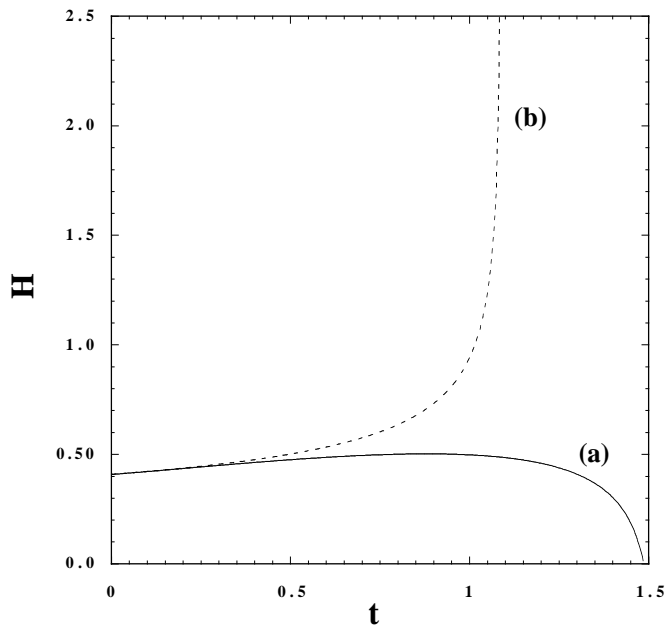}
\caption{The evolution of $H$ for the model
$f(\rho)=B\rho^{\beta}$ with $\beta=2$ and $B>0$.
The case (a) corresponds to the one in which quantum
corrections are taken into account with coefficients
$b=0.5$, $b'=-0.1$, and $b''=0$, whereas the case (b)
does not implement such effects. The Hubble rate approaches
$H=0$ with a finite time in the case (a), while it diverges
in the case (b).}
\label{Fig5}
\end{figure}

We numerically solve the background equations and
show the evolution of the Hubble rate
for $\beta=2$ and $B>0$ in Fig.~\ref{Fig5}.
In the presence of quantum corrections one has
$H \propto (t_s-t)^{1/3}$ around $t=t_s$, which means that
$H$ approaches zero. Meanwhile in the absence of quantum
corrections we have $H \propto (t_s-t)^{-1/3}$, thereby showing
the divergence of $H$ at $t=t_s$.
These properties are clearly seen in Fig.~\ref{Fig5}.

{}From Eq.~(\ref{EOSq4b}) we obtain
\be
\label{EOSq9}
a\sim a_0\exp\left[
\frac{\rho_0^{1-\beta}}{3B\left(1-\beta\right)}\left(t_s - t\right)
^{ \tilde\gamma\left(1 - \beta\right)}
\right]\,,
\ee
where $a_0$ is a constant.
Comparing the classical case [$\tilde{\gamma}=2/(1-2\beta)$]
with the quantum corrected one [$\tilde\gamma = 4/(1-2\beta)$],
we find that the power of $\left(t_s - t\right)$ is larger in the
presence of quantum corrections.
Then the scale factor approaches a constant $a_0$ more
rapidly if we account for the quantum effect, implying that the
spacetime tends to be smooth, although the divergence of $\rho$
is stronger. Thus quantum effects moderate the classical singularity.

\subsection{Type I singularity}

We shall next consider the model $f(\rho)\sim B\rho^{\beta}$
with $1/2<\beta<1$ when $\rho$ is large. In this case there exists the Big 
Rip singularity
as shown in Sec.~IV. We note that the classical evolution is
characterized by $\rho \propto (t_s-t)^{\frac{2}{1-2\beta}}$
and $H \propto (t_s-t)^{\frac{1}{1-2\beta}}$,
both of which exhibit divergence for $\beta>1/2$.

When the quantum correction is present, let us first assume that the 
time-dependence
of $\rho$ is given by Eq.~(\ref{EOSq4}) with negative $\tilde\gamma$.
Since $\tilde\gamma\left(1 - \beta\right)<0$ in this case,
we might expect that Eq.~(\ref{EOSq2}) would give the following approximate 
relation around $t=t_s$:
\be
\label{EOSq10}
\rho \sim 6b'H^4\,.
\ee
The term on the r.h.s. grows as
$H^4 \propto (t_s-t)^{-4+4\tilde\gamma (1-\beta)}$, but this does not give 
a
consistent result, since $\rho$ becomes negative for $b'<0$.

This tells that our assumptions should be wrong and $\rho$ does not become
infinite.
If $\rho$ has an extremum, Eq.~(\ref{EOSq3}) tells that 
$H$ vanishes there since $\dot \rho=0$.
In order to confirm whether the quantum effect moderates the singularity or 
not,
we have numerically solved the background equations for
the model where $\rho$ is exactly given by $f(\rho)=B\rho^{\beta}$ 
with $1/2<\beta<1$ and coefficients $b>0$, $b'<0$ and $b''=0$.
We find that the Hubble rate approaches zero with a
finite time, as is similar to the case (a) in Fig.~\ref{Fig5}.
Thus the presence of the quantum correction  moderates
the Big Rip singularity as well.

\subsection{Type II singularity}

Finally we consider the model $f(\rho)=C\left(\rho_0 - 
\rho\right)^{-\gamma}$
with $\gamma>0$ as an example of the sudden future type singularity.
By Eq.~(\ref{EOS35}) the energy density $\rho$ in the classical case
behaves as
\be
\label{EOSq13}
\rho \sim \rho_0 - \left\{\kappa C\sqrt{3\rho_0} (\gamma + 1) \left(t_s - 
t\right)
\right\}^{\frac{1}{\gamma + 1}}\,,
\ee
around the singularity at $t=t_s$.
Here we write $t_0$ in Eq.~(\ref{EOS35}) as $t_s$.
Using the Friedmann equation $H^2=\kappa^2 \rho/3$, we find
\be
\label{EOSq14}
H\sim \kappa\sqrt{\frac{\rho_0}{3}}\left\{1
- \frac{1}{2\rho_0}\left[\kappa C\sqrt{3\rho_0} (\gamma + 1) \left(t_s - 
t\right)
\right]^{\frac{1}{\gamma + 1}} 
\right\}\,.
\ee
Since $0<1/(\gamma + 1)<1$, $\dot H$ diverges at $t=t_s$ while $H$ is 
finite there.

Let us now include the quantum corrections.
We assume the following form of $\rho$ around $t=t_s$:
\be
\label{EOSq15}
\rho = \rho_0 + \rho_1 \left(t_s-t\right)^\nu\,,
\ee
with a positive constant $\nu$.
Using Eq.~(\ref{EOSq3}), one gets
\be
\label{EOSq16}
H\sim \frac{\nu\rho_1^{1+\gamma}}{3(-C)}
\left(t_s - t \right)^{-1 + \nu\left(1+\gamma\right)}\ .
\ee
Since $\nu\left(1+\gamma\right)>0$, we can use
the same approximate equation as in Eq.~(\ref{EOSq6}).
This gives
\bea
\label{EOSq18}
\nu = \frac{4}{2\gamma + 1}\,,
\eea
and
\bea
& & \rho = \rho_0 + \rho_1 \left(t_s-t\right)^{\frac{4}{2\gamma + 1}}\,, \\
& & H \propto (t_s-t)^{\frac{2\gamma+3}{2\gamma+1}}\,.
\eea

\begin{figure}
\includegraphics[height=3.3in,width=3.5in]{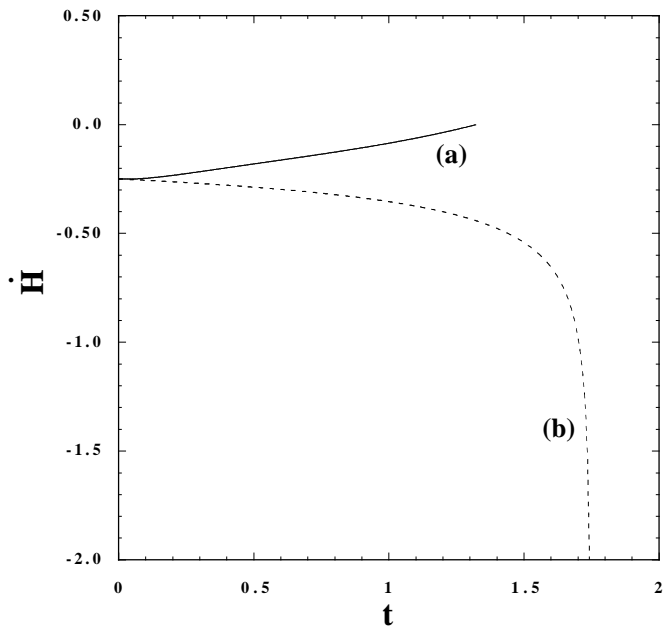}
\caption{The evolution of $\dot{H}$ for the model
$f(\rho)=C(\rho-\rho_0)^{-\gamma}$ with $\gamma=1/2$
and $C<0$.
We implement quantum effects in the case (a)
with coefficients $b=0.5$, $b'=-0.1$ and $b''=0$,
whereas the case (b) does not account for such effects.
We find that $\dot{H}$ approaches zero with a finite time
in the case (a), while it diverges in the case (b).}
\label{Fig6}
\end{figure}

Since $(2\gamma+3)/(2\gamma+1)$ is larger than 1,
not only $H$ but $\dot H$ are finite in the presence of quantum 
corrections.
This is numerically confirmed in Fig.~\ref{Fig6}.
Thus it is clear that quantum effects work to make the universe less 
singular
or completely non-singular (basically, asymptotically deSitter).
It was shown in Ref.~\cite{quantum} that this property also holds
for scalar-field dark energy models.

\vspace{1em}
Finally we should mention one point.
When the quantum correction becomes important, this typically
works to provide a negative energy density $\rho_A$ which nearly
cancels with the energy density $\rho$ of dark energy.
This is the reason why the Hubble rate does not diverge
if we account for the quantum effect.

\section{Discussion}

In summary, we discussed the fate of the (phantom) dark energy universe
by assuming the equation of state (EOS) with a form: $p=-\rho- f(\rho)$.
Our main interest is to clarify the structures of future singularities 
which
appear with finite time $t_s$.
We classified the types of singularities into four classes.
The type I is the Big Rip singularity \cite{CKW} at which
all of $a$, $\rho$ and $p$ exhibit divergences.
The type II corresponds to the sudden future
singularity in \cite{Barrow} at which $a$ and $\rho$ are finite
but $p$ diverges.
The type III is similar to the Big Rip, but the scale factor is finite.
The type IV is a mild singularity at which $a$, $\rho$, and $p$
are finite but higher derivatives of the Hubble rate diverge.
The model given in Eq.~(\ref{EOS15}) includes all of these singularities,
which are investigated in details in Sec.~III.
In Sec.~IV we adopted simpler forms of $f(\rho)$ which are obtained
by considering limiting cases of the function (\ref{EOS15})
and studied the general structure of singularities.
This is very helpful to understand when these
finite-time singularities appear.

In Sec.~II we wrote the background equations in an autonomous form
and showed that there exists one stable critical point if the constant
EOS ($w=p/\rho$) of dark energy is less than $-1$.
It was also found that there is no stable node for $w>-1$.
In Sec.~V we have numerically confirmed this property for the function
(\ref{EOS14}) that asymptotically gives constant values of $w$.
The singularity for $w<-1$ with the function (\ref{EOS14})
corresponds to the type I, which
implies that the Big Rip singularity is a late-time stable attractor.
We also performed phase-space analysis for the model (\ref{EOS15})
and found that the solutions approach ``instantaneous'' critical points
introduced in Sec.~II.
We have numerically checked that the late-time solutions agree well
with analytic estimations.

In Sec.~VI a (phantom) dark energy scenario coupled to
dark matter is investigated. We constructed the form of the
coupling $Q$ that reproduces the background evolution (\ref{EOS8})
giving the Big Rip singularity at $t=t_s$, see
Eqs.~(\ref{EOS37}) and (\ref{EOS41}).
Unlike the case of a single fluid, this allows a possibility to lead to
the transition from $w>-1$ to $w<-1$ without using a double-valued
function for $f(\rho)$.
We also explicitly presented a two-scalar field model
which gives the same singular behavior for the background evolution.

Finally, we accounted for the quantum correction coming from conformal
anomaly and studied its effect to type I, II and III singularities.
We found that this typically works to moderate the singularities or
prevent them
by providing a negative energy density.
In this case the divergent behavior of the Hubble rate
is prevented by the quantum effect, which is followed by the
decrease of $H$ toward 0.
Therefore the presence of the quantum effect can
drastically change the evolution of the universe.
For instance, the new inflationary era in far future
may be possible according to the conjecture of
Refs.~\cite{quantum,final}.

It is quite clear that with the growth of the energy density near to
singularity the effects of quantum gravity (string/M-theory) may become
dominant. Hence, it will be extremely important to estimate the effect of
dilatonic or modulus higher-order corrections \cite{corre} in dark energy 
models 
motivated by string theory, which we leave to future work.
{}From another point, even if stringy effects stop the future singularity
and possible phantom phase is transient, the growth of dark energy density
may give its imprints to current universe. It would be a challenge to
search for such imprints.

\section*{ACKNOWLEDGEMENTS}

We thank J.~Barrow for useful discussions.
This research has been supported in part by the Ministry of
Education, Science, Sports and Culture of Japan under
the grant n.13135208 (S.N.), RFBR grant 03-01-00105,
LRSS grant 1252.2003.2 and BFM2003-00620 grant (S.D.O).



\end{document}